# GMXPolymer: a generated polymerization algorithm based on GROMACS

Jianchuan Liu [a], Haiyan Lin [a], and Xun Li [b,*]

[a]School of Electrical Engineering and Electronic Information, Xihua University, Chengdu 610039, China

[b]Institute of Linguistics, Shanghai International Studies University, Shanghai, 200000, China

* Corresponding author. Email: xunli_heureux@163.com

**Abstract**

This work introduces a method for generating generalized structures of amorphous polymers using simulated polymerization and molecular dynamics equilibration, with a particular focus on amorphous polymers. The techniques and algorithms used in this method are described in the main text, and example input scripts are provided for the GMXPolymer code, which is based on the GROMACS molecular dynamics package. To demonstrate the efficacy of our method, we apply it to different glassy polymers exhibiting varying degrees of functionality, polarity, and rigidity. The reliability of the method is validated by comparing simulation results to experimental data for various structural and thermal properties, all of which show excellent agreement.

**Keywords:** Polymerization, GROMACS, MD simulation, Bonds, Polymer

## 1 Introduction

Glassy polymers, also known as amorphous polymers, are a class of materials that exhibit unique mechanical, thermal, and electrical properties due to their disordered molecular structure [1-10]. Unlike crystalline polymers, glassy polymers lack long-range order and instead have a highly entangled and disordered molecular structure. Polymers' glassy state is achieved by rapidly cooling the melt or using suitable solvents that prevent crystallization during the polymerization process. Researches on glassy polymers have gained significant attention recently due to their potential applications in various industries, including packaging, electronics, and aerospace[11-15]. One of the major challenges in studying glassy polymers is to understand their structure-property relationships, which are highly dependent on their molecular structure and processing conditions.

Experimental studies of glassy polymers have provided important insights into their behavior and properties. These results have enabled the development of structure-property relationships for these materials. They have contributed to developing new materials with tailored properties for various applications[16-23]. Several studies have investigated the effect of molecular weight, molecular architecture, and processing conditions on the glass transition temperature (Tg) and other properties of glassy polymers[24-28]. In addition, experimental techniques have been developed to probe the behavior and properties of glassy polymers, including mechanical testing[1, 29-32], thermal analysis[1, 33-36], and spectroscopy[5, 37, 38]. For example, Wei et al.[39] used dynamic mechanical analysis to investigate the effect of molecular weight and cooling rate on the glass transition behavior of polystyrene. Lee et al.[40] used differential scanning calorimetry and thermogravimetric analysis to investigate the effect of molecular weight and cooling rate on the glass transition behavior and poly(methyl methacrylate) thermal stability. Wu et al.[41] studied the effect of molecular weight, molecular architecture, and processing conditions on glassy polymers' structure and dynamics using infrared or nuclear magnetic resonance spectroscopy. In conclusion, experimental studies of glassy polymers have provided valuable insights into the behavior and properties of these materials.

However, the complex molecular structure and processing conditions of glassy polymers make it difficult to understand their behavior at the molecular level in experiments. Molecular dynamics (MD) simulations have emerged as a powerful tool for investigating the behavior of glassy polymers. MD simulations can provide information on the local and global structures, thermodynamics, and transport properties of glassy polymers at the molecular level. MD simulations also allow for the systematic investigation of the effect of various factors on the behavior of glassy polymers, such as molecular weight, temperature, molecular architecture, and processing conditions. The recent progress in MD simulations of glassy polymers has provided valuable insights into the relationship between molecular structure and dynamics, glass transition behavior, mechanical properties, and thermal stability of these materials [42-53]. For example, Li et al. [53] utilized molecular dynamics simulations to investigate the effect of molecular weight and temperature on the glassy behavior of polystyrene. Nguyen et al. [43] investigated the influence of molecular architecture on the glass transition temperature of polymer thin films using a combination of experimental and theoretical methods. These studies have contributed significantly to understanding the structure-property relationships in glassy polymers.

However, challenges and limitations still exist, and further research is needed to overcome these challenges and to fully exploit the potential of MD simulations in the study of glassy polymers. In this process, a major challenge is about the accuracy of molecular structures and simulation force field files, which are used to describe the interactions between the atoms or molecules in the system. Inaccurate molecular structures and simulation force field files can lead to errors in the simulation results, particularly for glassy polymers. Generating accurate molecular structures and simulation force field files for amorphous polymers is difficult for several reasons. Firstly, amorphous polymers lack a well-defined repeating unit, making it hard to generate a molecular structure that accurately represents the bulk material. Secondly, amorphous polymers typically have a high glass transition temperature (Tg), which requires an accurate representation of the intermolecular interactions between polymer chains. Finally, the structure of amorphous polymers is highly dependent on processing conditions such as temperature and pressure, making it difficult to simulate a representative structure.

Researchers have developed various methodologies for generating molecular structures and simulation force fields for amorphous polymers. These methodologies typically combine experimental data with computational simulations to generate a representative molecular structure and simulation force field. For example, the ReaxFF [54] force field has been used to simulate the polymerization of amorphous polymers, which allows for the prediction of the structural and thermal properties of the material. Similarly, the Polymer Reference Interaction Site Model (PRISM)[55] has been used to generate a realistic representation of the intermolecular interactions between polymer chains. Besides, numerous techniques have been published in the literature to generate structures for amorphous materials of glassy polymers [56, 57]. However, many of these efforts are custom-tailored for specific materials and circumstances, and specific algorithms are not consistently provided, resulting in a lack of comprehensive understanding of the methods employed. Despite these advances, further research is still needed to develop more accurate and efficient methodologies.

It's worth noting that Colina et al.[58] utilized a simulated polymerization technique coupled with a 21-step molecular dynamics equilibration process. This method enables the generation of structures that are not limited to specific polymer types, allowing for a universal application. Additionally, the use of molecular dynamics equilibration aids in ensuring that the resulting structure is thermodynamically stable and resembles a physically realistic amorphous polymer structure. However, this approach is exclusive to the LAMMPS program[59] and lacks compatibility with the GROMACS program[60], which is widely utilized. Consequently, numerous users relying on the GROMACS program for polymer modeling and simulation are faced with a dearth of efficacious methods to carry out their endeavors.

In this study, we present an universal polymerization algorithm predicated upon the GROMACS program's utilization, with an explicit description of its use as applied with the GMXPolymer code[61]. This approach is user-friendly in adjusting simulation parameters. Users can tailor the MDP file of the GROMACS program to their specific research system.

## 2 Method details

### 2.1 algorithm

The basic structure of the simulated polymerization algorithm, as utilized in this work, is as follows and shown Fig.1:

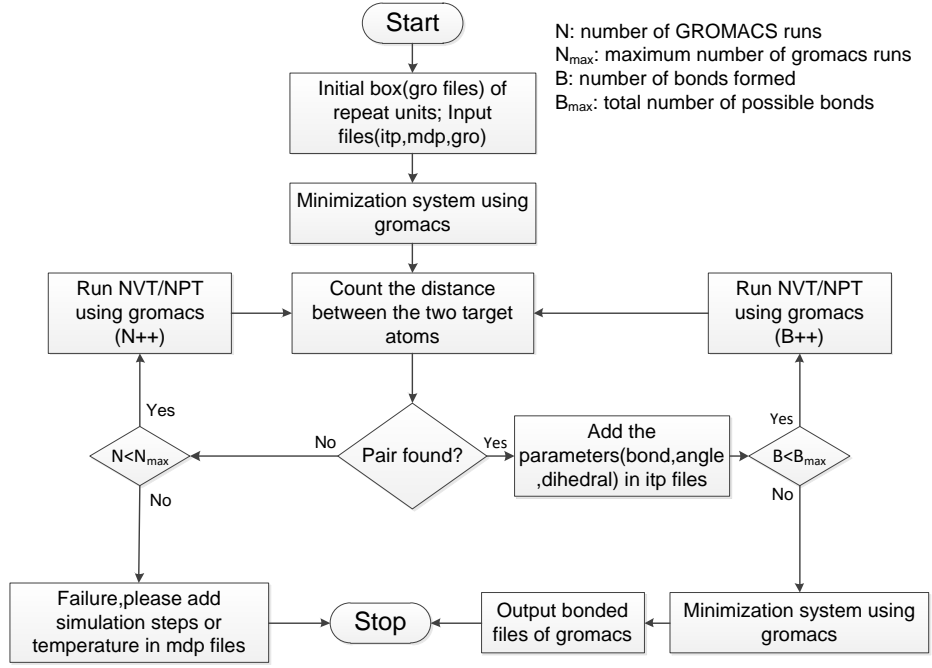

**Fig. 1** Flow chart of the GMXPolymer code

1. Initialization: producing a simulation box of GROMACS containing specific proportion monomer; preparing ITP files of GROMACS program for each monomer; preparing MDP files of GROMACS program for EM and NPT or NVT; preparing the DAT file containing parameters of the new bond, angles, dihedrals.

2. A polymerization step is performed: (a) running the GROMACS program to relaxation structure; (b) calculating the distances of all target bonding atoms; (c) the pair of reactive atoms with the minimum distance that is less than cut distance is selected. Generated a new ITP file and added parameters of new bond, angles, and dihedrals; (d) If the distance of pair of reactive atoms is not less than the cut distance, a molecular dynamics simulation is performed, and a polymerization step is attempted again. This is repeated up to $N_{max}$ times until a bond is formed. It is noteworthy that the newly added parameters of bond, angle, and dihedrals follow the formula below:

$$E_b = K_b(r - r_0)^2 \qquad (1)$$

$$E_a = K_a(\theta - \theta_0)^2 \qquad (2)$$

$$E_d = \frac{V_1}{2}[1 + cos\phi] + \frac{V_2}{2}[1 - cos2\phi] + \frac{V_3}{2}[1 + cos3\phi] +$$
$$\frac{V_4}{2}[1 - cos4\phi] + \frac{V_5}{2}[1 + cos5\phi] + \frac{V_6}{2}[1 - cos6\phi] \qquad (3)$$

where, $E_b$, $E_a$ and $E_d$ are bond energy, angle energy, and dihedral energy, respectively. $K_b$ and $K_a$ are a force constant. $r$ is the distance between the two atoms considered, and $r_0$ is the equilibrium bond distance. $\theta$ is the bond angle, and $\theta_0$ is the equilibrium angle. $V_1$, $V_2$, $V_3$, $V_4$, $V_5$ and $V_6$ are coefficients in the Fourier series having units of energy and $\emptyset$ is the dihedral angle.

3. Cycles of polymerization steps are successively repeated until $B_{max}$ bonds are formed or until no pair meeting the bonding criteria is identified within the $N_{max}$ molecular dynamics simulations. At the end of the $B$ cycle, a short minimize molecular dynamics step is carried out to relax any remaining stresses in the system and allow structural rearrangement of the configuration.

It should be noted that if the cycle reaches the maximum number of times, the required number of bonds is not obtained, or there is a running error in the process of calling the GROMACS program, it should increase the simulation time and increase the simulation temperature in the MDP file so that it can be obtained more sampling configurations and elimination of unreasonable configurations. In addition, when the polymer system is successfully formed, it is recommended to use GROMACS to perform the high-temperature, high-pressure, and annealing simulation. The high-temperature and high-pressure steps allow large energy barriers to be overcome to speed up the normally lengthy relaxation of polymers.

**2.2 the engine runs the code**

This work implements the GMXPolymer simulated polymerization algorithm on GROMACS program [62]. GMXPolymer code controls the main polymerization loop. The energy minimizations and molecular dynamics simulations use the GROMACS program called by the GMXPolymer code. A new ITP file is generated when a new bond is formed, and the necessary additions to the ITP file are made to include new bonds, angles, and dihedrals. In preparing the ITP file of the monomer, the charge of the reactive atom must be modified before the code runs so that it is a correct value after bonding.

## 3 Results and discussion

To illustrate the generality of the structure generation techniques described above, different polymers (Fig. 1) spanning methods were studied. All the MD simulations were performed using the GROMACS-2023.2 software package[62] in this code. For all the systems, the intra- and inter-molecular interaction parameters were built from the general AMBER force field (GAFF)[63]. The partial charges were obtained by the restrained electrostatic potential (RESP) fitting method [64, 65]. A spherical cutoff of 1.2 nm for the summation of van der Waals (VDW) interactions and the particle-mesh Ewald (PME) solver for long-range Coulomb interactions with a cutoff of 1.2 nm and a tolerance of $1.0 \times 10^{-5}$ were used throughout. The simulations were carried out using the leap-frog integrator with a time step of 1 fs. The Nose-Hoover thermostat[66] and Parrinello-Rahman barostat[67] were applied for the temperature and pressure control.

The GMXPolymer code was applied to each in the same manner. The repeat units were specifically designed to provide the proper polymerized structures by adding one (or more) bonds and not deleting atoms. When preparing the monomer structure file, it is necessary to deal with the connection atoms, that is, delete the edge link atoms not in the polymer, as shown in Fig. 2. The polymerization steps are illustrated in Fig. 3 between thiophene monomer and between trimesoyl chloride (TMC) and 5,5',6,6'-Tetrahydroxy-3,3,3',3'-tetramethyl-1,1'-spirobiindane (TTSBI) monomer, respectively. Following the crosslinking of thiophene monomers, new C-C bonds, S-C-C angles, and dihedral angles like S-C-C-S and S-C-C-C are formed. Similarly, in the polymerization of TMC and TTSBI monomers, new C-O bonds, C-O-C angles, as well as dihedral angles like C-O-C-O and C-O-C-C were created. These bonding parameters need to be pre-defined in the "bond.dat" file, after which the program automatically identified these bonding details and incorporates them into the ITP file post crosslinking. Therefore, the accuracy of simulation results is contingent upon the user's pre-specified bonding information.

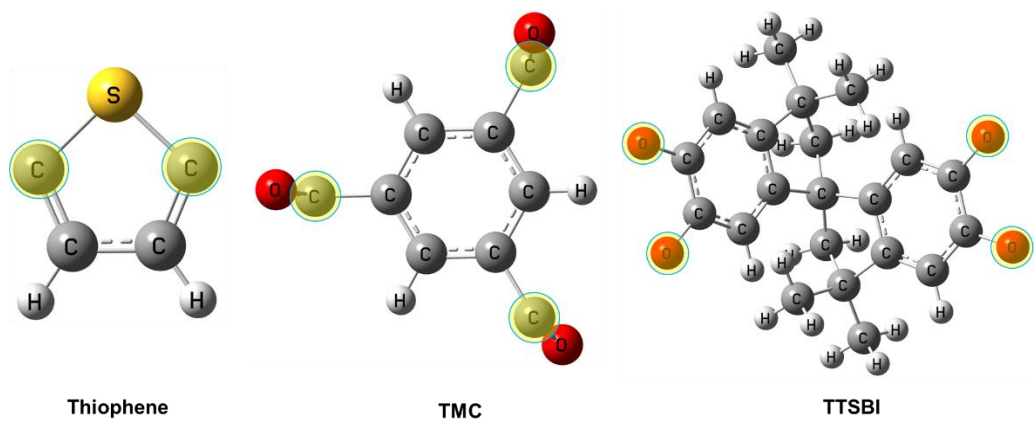

**Figure 2.** Chemical structures of pre-cross-linked monomer: thiophene, trimesoyl chloride (TMC), 5,5',6,6'-Tetrahydroxy-3,3,3',3'-tetramethyl-1,1'-spirobiindane (TTSBI). The reaction atoms are indicated by the blue circular markers.

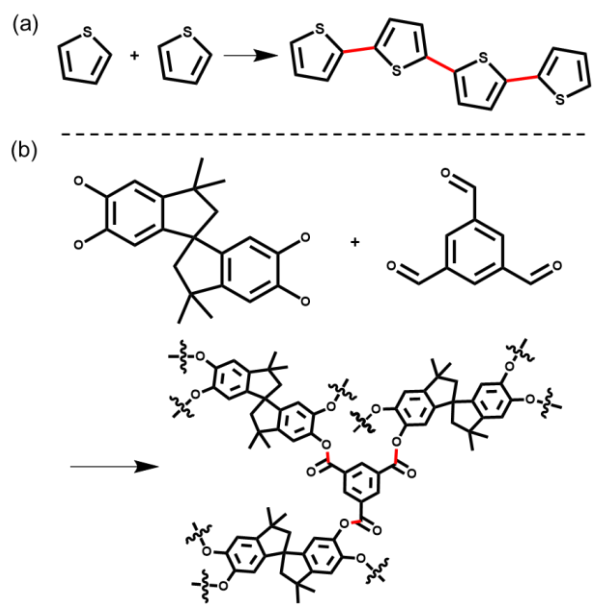

**Figure 3.** Polymerization steps illustrated (a) between thiophene monomer and (b) between TMC and TTSBI monomer, respectively.

As illustrated in Fig. 4, crosslinked structures containing different numbers of thiophene monomers were constructed. Whenever the distance between target reactive atoms (see Fig. 2) on different thiophene monomers fell below the specified cutoff radius (set to 6 Å in this study), bonding information was added, encompassing bond types, angles, dihedral angles, and related details. As can be seen in Fig. 4, it is evident that due to strong van der Waals repulsion, atoms couldn't approach closely during the simulation, resulting in initially unrealistic long bond lengths. Therefore, before proceeding, newly formed bonds were rapidly relaxed through energy minimization to achieve structures closer to reality. It is important to note that only the closest pair of atoms was chosen for bonding each time, followed by running GROMACS for short-step molecular dynamics simulations and subsequent recalculations of distances between all target reactive atoms. Additionally, our program allows for the specification of intra-molecular crosslinking within polymer chains. For polymers containing thiophene monomers, we constrained internal molecular crosslinking to encourage the formation of more realistic linear structures.

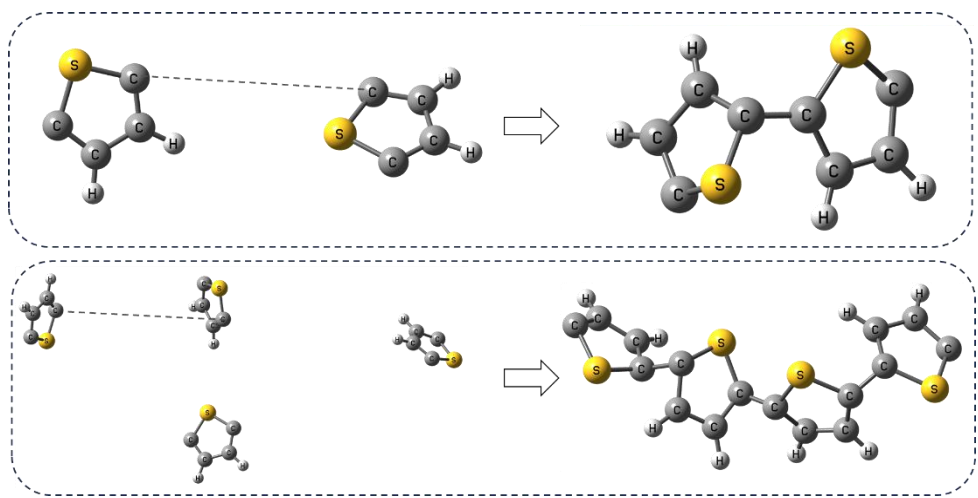

**Figure 4.** Cross-linked simulation of thiophene monomers with different numbers.

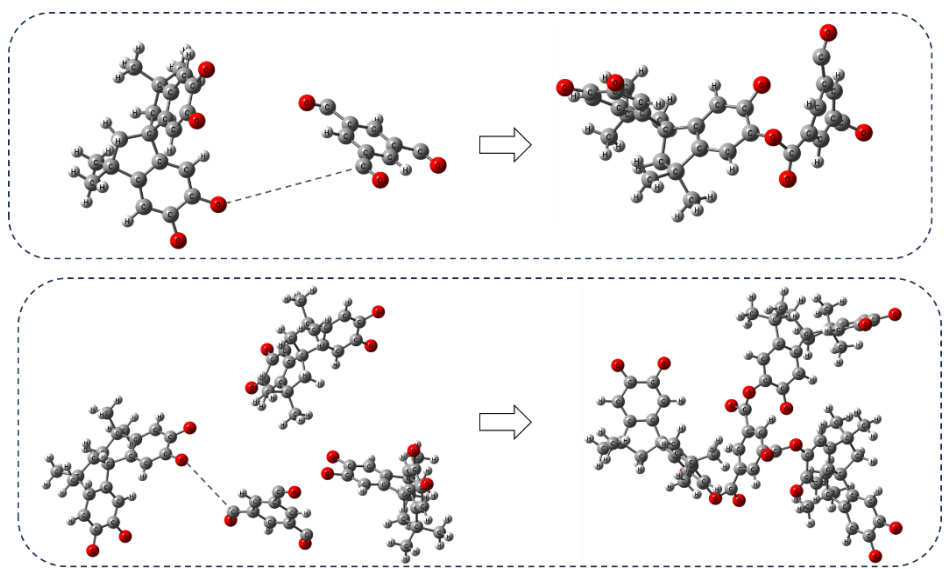

**Figure 5.** Cross-linked simulation of between TMC and TTSBI monomers with different numbers.

Similarly, the program also supports intermolecular polymerization crosslinking and multi-site crosslinking. We established crosslinking systems comprising varying numbers of TTSBI monomers and TMC molecules. Within the TMC monomers, three reactive sites were configured (as shown in Fi. 2), while the TTSBI molecules were set with four reactive sites (as depicted in Fig. 2). As shown in Fig. 5, reasonable structural configurations were obtained post crosslinking reactions. Fig. 6 illustrated the polymer structure formed by the polymerization of 50 TMC monomers and 75 TTSBI monomers. In this case, we permitted intra-molecular bonding crosslinking, contributing to properties consistent with those of glassy polymers.

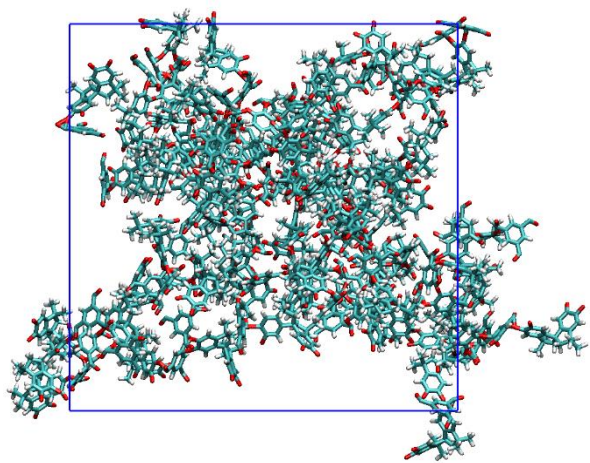

**Figure 6.** Polymer molecular structure of cross-linked between TMC and TTSBI.

The importance of poly (methyl methacrylate) in the biomedical field is evident and the crosslinked forms of poly (methyl methacrylate) had been utilized in many applications [68-70]. In recent years, researchers have conducted in-depth studies on the relationship among the crosslinking density, polymer network structure, and dynamic mechanical properties of various crosslinked poly(methyl methacrylate) [68-71]. The main polymerization principle involves a crosslinking monomer molecule with a double bond reacting with a crosslinking agent molecule with two double bonds to form a crosslinked structure, as shown in Fig.7(a). These crosslinked network molecular structures, like the chemical structure, determine the physical and mechanical properties of thermosetting resins[72]. Therefore, the analysis and understanding of the structure and properties of crosslinked poly(methyl methacrylate) are of scientific and technological importance in the biomedical field. In this paper, a series of model poly(methyl methacrylate) networks were crosslinked by the introduction of: 1, 2, 5, 10 and 20 mol.% of triethylene glycol dimethacrylate (TEGDMA) by the GMXPolymer code.

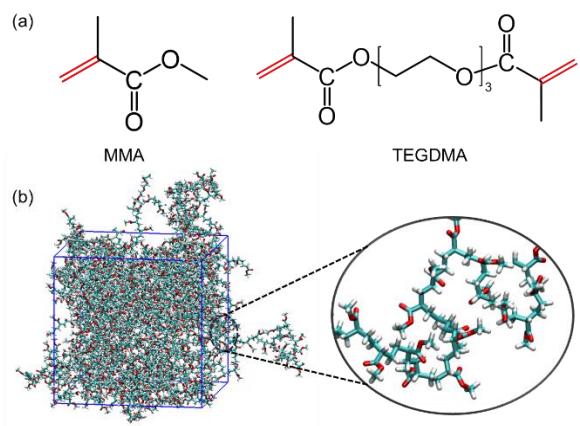

**Figure 7.** (a) the chemical structure of the MMA and TEGDMA monomers. (b) polymer molecular structure of cross-linked between MMA and TEGDMA monomers and the cross-linking structure local magnification configuration.

Five polymer networks with varying crosslink densities were constructed via the thermal copolymerization in bulk of methyl methacrylate (MMA) with triethylene glycol dimethacrylate (TEGDMA). Fig.7(b) illustrates the simulated polymer network for the 1 mol.% of TEGDMA, and the cross-linking structure local magnification configuration. After crosslinking, we performed 5 ns MD simulations of the five systems at 300 K and 1 atm using the NPT ensemble to calculate their densities, shown in Table 1. From Table 1, it can be observed that as the molar ratio of TEGDMA increases, the density of the cross-linked system also increases. This trend is consistent with both our simulation results and the experimental data[71].

**Table 1** the densities ($d$, g/cm$^3$) for the 1, 2, 5, 10 and 20 mol.% of TEGDMA.

| | TEGDMA mol.% fraction | | | | |
|---|---|---|---|---|---|
| | 1 | 2 | 5 | 10 | 20 |
| $d$ (Simu.) | 1.11 | 1.12 | 1.12 | 1.13 | 1.15 |
| $d$ (Exp.[71]) | 1.17 | 1.18 | 1.18 | 1.19 | 1.20 |

The glass transition temperature ($T_g$) has a significant impact on the performance, processing, operating temperature range, and stability of polymer materials, making it one of the key parameters to consider when designing and selecting polymer materials.

To estimate $T_g$, we performed thermal quenching simulations while monitored the density to obtain the thermal curves and the result was shown in Fig.8(a). The glass-transition region was defined by the limited temperature range in which the thermal expansion coefficient experienced a change, and $T_g$ was estimated as the point where the linear fits to the glassy and melt regions intersected[73]. Fig.8(b) presented the comparison between simulation results and experimental data. It can be found that 20 mol.% of TEGDMA had the highest $T_g$, and the simulation values is slightly larger than the experimental results[71]. Taking into consideration both experimental data and computational results, we conclude that the GMXPolymer code can serve as an effective tool for studying the cross-linking reactions of the target system.

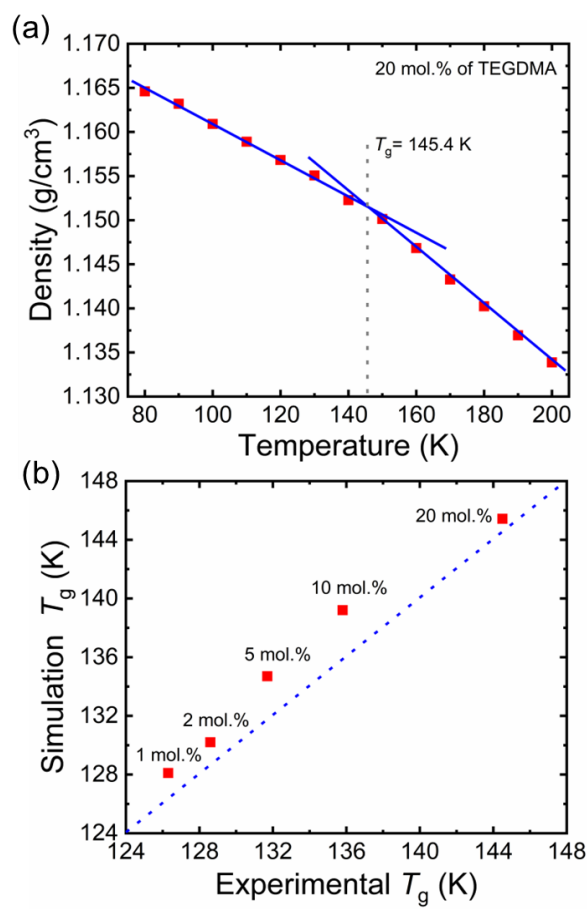

**Figure 8.** (a) The density vs. temperature curves showing glass-transitions of 20 mol.% of TEGDMA. (b) Simulation $T_g$ of mol.% of TEGDMA and available experimental[71].

## 4 Conclusions

In conclusion, this study presents an approach for generating generalized structures of amorphous polymers through simulated polymerization and molecular dynamics equilibration, focusing specifically on amorphous polymer systems. The GMXPolymer code was systematically applied, designed to create polymerized structures by selectively adding bonds without deleting atoms, enabling the generation of proper repeat units. Preparation of the monomer structure file involved managing connection atoms, specifically deleting edge link atoms not within the polymer. The GMXPolymer code allows users to specify intra-molecular crosslinking, steering the formation of more realistic linear structures within polymers. Moreover, the program supports intermolecular polymerization crosslinking and multi-site crosslinking, exemplified by crosslinking systems constructed from diverse TTSBI monomers and TMC molecules, or methyl methacrylate (MMA) and triethylene glycol dimethacrylate (TEGDMA) molecules. Permitting intra-molecular bonding crosslinking in these systems produced properties consistent with those of glassy polymers, affirming the versatility and utility of our approach in modeling diverse amorphous polymer structures.

**Acknowledgments**

The authors thank Dr. Shuang Liu for his input regarding the cross-linked procedure.

**Appendix: Installation guide**

The following instructions pertain to the download and installation of the latest version of GMXPolymer. Download the latest version of GMXPolymer from: https://github.com/lauthirteen/GMXPolymer